\documentstyle[preprint,eqsecnum,aps]{revtex}
\begin{document}
\draft
\preprint{\vbox{\baselineskip=12pt
\rightline{
CGPG-93/8-5, \
hep-th/9308150}}}
\title{The Generalized Peierls Bracket}
\author{Donald Marolf\cite{Marolf}}
\address{Physics Department, The Pennsylvania State University,
University Park, PA 16802} \date{revised February, 1994}
\maketitle

\begin{abstract}
We first extend the Peierls algebra of gauge invariant
functions from the space ${\cal S}$ of classical solutions
to the space ${\cal H}$ of histories used in path integration and
some studies of decoherence.  We then show that it
may be generalized in a number of ways to act on gauge
dependent functions on ${\cal H}$.  These generalizations
(referred to as class I)
depend on the choice of an ``invariance breaking term,"
which must be chosen carefully so that the gauge dependent
algebra is a Lie algebra.  Another class of invariance breaking
terms is also found that leads to an algebra of gauge dependent functions, but
only on the space ${\cal S}$ of solutions.
By the proper choice of invariance breaking term,
we can construct a generalized Peierls algebra that agrees
with any gauge dependent algebra constructed through
canonical or gauge fixing  methods, as well as Feynman and
Landau ``gauge."  Thus, generalized Peierls algebras
present a unified description of these techniques. We
study the properties of generalized Peierls  algebras and
their pull backs to spaces of partial solutions and find
that they may posses constraints similar to the canonical
case. Such constraints are always first class, and
quantization may proceed accordingly.

\end{abstract}
\pacs{}

\section{Introduction}
\label{intro}

In \cite{other}, we began an investigation of classical
(commuting) *-Lie algebras ${\cal A}({\cal H})$ on spaces
${\cal H}$ of histories. These algebras contained smooth
complex functions on ${\cal H}$ with the usual operations
of multiplication, addition, and complex conjugation (*)
as well a Lie bracket operation defined
by extending the Poisson bracket to ${\cal H}$ from the
phase space $\Gamma$ or  reduced phase space $\Gamma_r$.
This Lie bracket was often assembled from pieces defined
locally.

Our motivation here as in \cite{other} is to define
algebras on ${\cal H}$, the space ${\cal S}$ of solutions,
and the space ${\cal E}$ of evolutions (introduced in
\cite{other}) from which quantum theories can be
derived.  The hope is that such
algebras will lead to a better understanding of the relationship
between path integral and algebraic methods, provide a
unified description of conventional gauge dependent algebras,
and suggest new avenues for quantization.  This
approach is complementary to that of \cite{Wald},
\cite{Abhay}, \cite{Witten} and others which define a
presymplectic structure on ${\cal H}$ and ${\cal S}$ since
this presymplectic form and (often) our Lie algebras are
degenerate so that neither can be inverted to obtain
the other.

The close relationship between this extension and the
usual Poisson bracket allowed us to compare the extended
Poisson algebra ${\cal A}_H({\cal H})$ with the familiar
Poisson algebras ${\cal A}_H(\Gamma)$ and ${\cal
A}_H(\Gamma_r)$ on the phase space $\Gamma$ and reduced
phase space $\Gamma_r$ with little effort and to study
quantization of ${\cal A}_H({\cal H})$, ${\cal A}_H({\cal
E})$, and ${\cal A}_H({\cal S})$ where ${\cal A}_H({\cal
E})$, and ${\cal A}_H({\cal S})$ are pull backs of ${\cal
A}_H({\cal H})$ to ${\cal E}$ and ${\cal S}$.  We saw that
each of these algebras leads to a Heisenberg picture
quantization resembling Dirac's constraint quantization
\cite{Dirac} but with certain differences. We also saw how
canonical and gauge fixed Poisson algebras are both
examples of ``gauge breaking" schemes.

Here, we develop a more general construction of classical
*-Lie algebras on ${\cal S}$ and ${\cal H}$ by extending
and generalizing the Peierls bracket\cite{Peierls}.  This
construction, introduced in \cite{Dis}, uses more
machinery than that of \cite{other} but provides a unified
perspective and its covariance is manifest. The work in
\cite{other} is an important link between
the material presented here and more familiar techniques and it
is strongly recommended that \cite{other} be read before
studying what follows.  The extended
Poisson bracket will be shown to be a special case of the
generalized Peierls bracket so that a comparison of these
general methods with ${\cal A}_H(\Gamma)$, ${\cal
A}_H(\Gamma_r)$, and the usual quantization techniques
follows from \cite{other}.

We begin in section \ref{Ext} with an introduction of
notation as well as a brief
review of the Peierls algebra ${\cal A}_L({\cal S})$
and the associated techniques of \cite{Bryce}
which may be unfamiliar.  The {\it extension} ${\cal
A}_L({\cal H})$ of the Peierls bracket to ${\cal H}$ for
gauge-free systems is then direct.  Appendix \ref{equiv}
shows that this extension is equivalent to the extended
Dirac algebra of \cite{other}.

Using additional techniques of \cite{Bryce},
section \ref{Gen} discusses the case of gauge systems and
describes the {\it generalization} of the Peierls bracket
to act on gauge dependent functions.  A subtle point
concerning the spacetime support of these functions is
discussed in appendix \ref{boundaries}.  This
generalization depends on the choice of an ``invariance
breaking term" which must be chosen  carefully so that the
generalized Peierls algebra is a Lie algebra.  The
difficult property to ensure is the Jacobi identity and
Appendix  \ref{fail} gives two examples for which this
identity fails to hold.   However, section \ref{Gen} finds two
general classes of invariance breaking terms guaranteed to
produce Lie algebras, one of which defines a Lie bracket on
${\cal A}_L({\cal H})$ and one of which defines such a
bracket only on ${\cal A}_L({\cal S})$.

Section
\ref{explore} explores the range of this generalization.
With the help of Appendix \ref{equiv}, it shows that the
generalized Peierls algebra includes the extended Poisson
algebras of \cite{other}.  In particular, both the
canonical algebra  of \cite{Dirac} and gauge fixed
algebras can be derived from our procedure.  Section
\ref{explore} also shows that generalized
Peierls algebras include the Landau and Feynman ``gauge"
algebras, so that we have found a unified description of
all of these techniques.

Our concentration on ${\cal H}$ provides much of this
unification since ${\cal S}$, ${\cal E}$, and other
interesting spaces are contained in ${\cal H}$.  However,
it is often desirable to deal with these subspaces
directly.  Because we consider ${\cal E}$ and ${\cal S}$
as subspaces of ${\cal H}$ and not as projections as in
\cite{Wald}, the natural notion is that of a pull back of
${\cal A}_L({\cal H})$.  Section  \ref{pull backs}
identifies subspaces to which such pull backs are
well-defined and investigates their properties.  The
resulting algebras, as well as ${\cal A}_L({\cal H})$
itself, posses ``generalized constraints" which
are analogous to the constraints of the canonical theory
but which are always first class.  Appendix \ref{solving}
shows how some pull backs may be obtained directly as
generalized Peierls algebras on smaller  spaces of
histories.  A summary discussion appears in section \ref{Dis}.

\section{Extending the Peierls Bracket}

\label{Ext}

In this section, we describe an extension of the usual
Peierls bracket to a space ${\cal H}$ of histories
as preparation for our later {\it generalization} of
the Peierls bracket to gauge dependent functions.
This ${\cal H}$ is to be the domain of some action
functional $S$ that is stationary on the space ${\cal S}
\subset {\cal H}$ of solutions.  Typically, ${\cal H}$
will be the space of sufficiently  regular fields on some
differential manifold $M$.   We note that the space ${\cal
H}$ is not an inherent property of the system, but depends
on our description through the choice of action $S$.  For
example, the space of histories for a scalar field depends
on whether we use a canonical or covariant description.

The underlying manifold $M$ may or may not have some
associated background structure such as a Lorentzian
metric or causal structure. We will, however, make use of
its differential structure and consider a number of
distributions that are local or ultralocal on $M$. We also
assume that variations $S,_i$ of $S$ with respect to
coordinates $\phi^i$ for $i \in {\cal I}$ on ${\cal H}$
yield a set of differential  equations on $M$ that has a
causal structure which may be used to define Cauchy surfaces in $M$,
though this structure may be dynamically determined through
the fields $\phi^i$.

We will usually take the coordinates $\phi^i$ to refer to
the values of fields at points in spacetime.  Thus, the
index $i$ contains a label for the field and for the
spacetime point and $\phi^i$ is ultralocal in $M$.
However, we assume only that such coordinates exist locally
so that ${\cal H}$ is an infinite dimensional manifold.
Certain steps may require that ${\cal H}$ be given further
topological properties but we will not address this level
of technicality.  Note that we use the condensed notation
of \cite{Bryce} as well as abstract index notation.

Our extension $(,)_{\cal H}$ is to be a Lie bracket of
functions on ${\cal H}$.  This means that it must be
bilinear and antisymmetric and must satisfy the Jacobi
identity and the derivation requirement: \begin{equation}
\label{derivationreq}
(AB,C) = A(B,C) + (A,C)B
\end{equation}
It follows that the bracket is determined by the
antisymmetric contravariant tensor field $\tilde{G}^{ij}
= (\phi^i,\phi^j)$ on ${\cal H}$ through $(A,B) =
A,_i \tilde{G}^{ij} B,_j$.

Because the Peierls bracket is unfamiliar to many
researchers, subsection \ref{background} presents a brief
review. Subsection \ref{ext} then defines $(,)_{\cal H}$
for gauge-free systems using the techniques of
\cite{Bryce}.

\subsection{The Peierls Bracket}
\label{background}

In 1952, R.E. Peierls noticed that an algebraic structure
equivalent to the Poisson bracket could be defined
directly from any action principle without first
performing a canonical decomposition into coordinates and
momenta.  His essential insight was to consider the
advanced and retarded ``effect of one quantity ($A$) on
another ($B$)."
Here, $A$ and $B$ are to be
functions on ${\cal H}$.  They
may be nonlocal in $M$, but
only in such a way that there
exist cauchy surfaces both to the future and to the past
of the support of $A$ and $B$. More general cases may be
defined through limits when those limits converge.

The advanced ($D^+_AB$) and retarded ($D^-_AB$) effects of
$A$ on $B$ are then defined by comparing the original
system with a new system defined by the action
$S_{\epsilon} = S + \epsilon A$ and the same space ${\cal
H}$ of histories.  Under retarded (advanced) boundary
conditions for which  the solutions $\phi^i \in {\cal S}$
and $\phi^i_{\epsilon} \in  {\cal S}_{\epsilon}$ coincide
to the past (future) of the support of $A$, the quantity
$B_0 = B(\phi^i)$ computed using $\phi^i$ will in general
differ from $B_{\epsilon} = B(\phi^i_{\epsilon})$ computed
using $\phi^i_{\epsilon}$.  For small epsilon, the
difference between these quantities defines the retarded
(advanced) effect of $A$ on $B$ through:
\begin{equation}
\label{effects}
D^{\pm}_AB = lim_{\epsilon \rightarrow 0}
\case{1}{\epsilon}(B_{\epsilon}-B_0
)\end{equation}
which depends on the unperturbed solution
$\phi^i$.

The Peierls bracket is defined to be the difference of
these two quantities:
\begin{equation}
\label{Peierls}
(A,B) = D^+_AB - D^-_AB = D^-_BA-D^-_AB
\end{equation}
where the last equality follows from the fact that the
dynamics is described by an action principle (see
\cite{Peierls}, but we will discuss a similar issue in
section \ref{ext}).  The Peierls bracket is in fact a Lie
bracket; the antisymmetry of this bracket is clear from
the last line above and Peierls \cite{Peierls}  shows that
\ref{Peierls} satisfies the Jacobi identity and the
derivation requirement.

Note that the Peierls bracket is defined on functions $A$
and $B$  of {\it solutions}; functions on the subspace
${\cal S}$ instead of all of ${\cal H}$.  For a system
without constraints, any map that evaluates the phase
space coordinates at some time $t$ takes ${\cal S}$ to the
phase space $\Gamma$ and Peierls' original paper
\cite{Peierls} shows that this map is a Lie bracket
isomorphism between the Poisson and Peierls algebras.

Our main interest, however, will be in systems with gauge
symmetries. For such a case, the Peierls bracket
can be defined as above only when
$A$ and $B$ are gauge invariant quantities.  If, say, $B$
is gauge dependent, then the value of
$B$ will not be uniquely determined by the initial data
and $D^-_AB$ will be ill-defined.  Recall that it was
necessary to define both $D^-_AB$ and $D^-_BA$ to verify
antisymmetry of the Peierls  bracket using \ref{Peierls}.
The above argument breaks  down if $A$ or $B$ is gauge
dependent.

If we are to define a Peierls bracket of gauge dependent
functions, it appears that we need a sort of gauge fixing
for the disturbances $\delta  \phi^i = \phi^i_{\epsilon}
-\phi^i$ which requires a  local gauge fixing in ${\cal
S}$. The structure we will introduce in section \ref{Gen}
is more general than this but is based on techniques of
\cite{Bryce} that relate to such a gauge fixing.  As such,
the formalism and notation of \cite{Bryce} will be convenient.
In \ref{ext}, we introduce this formalism in the gauge-free case and
find that it leads directly to the desired extension $(,)_{\cal H}$.

\subsection{Extensions to the Space of Histories}
\label{ext}

This section represents a brief
digression from the main line of discussion.  Instead of
immediately generalizing the Peierls bracket to gauge
dependent functions, we first extend it to the space of
histories ${\cal H}$. This extension will prove useful in
our study of the generalized Peierls algebra defined in
\ref{QFs} and provides the opportunities to introduce
notation and techniques. We first define the advanced and
retarded effects of $A$ and $B$ on each other as functions
on ${\cal H}$, from which the extended the Peierls bracket
follows. This will be straightforward using the machinery
of \cite{Bryce} and indeed, much of what follows is
implicit in that  treatment.

Following \cite{Bryce}, recall that the undisturbed fields
satisfy the equations of motion:
\begin{equation}
\label{eom}
0 = S,_i(\phi^j)
\end{equation}
while the disturbed fields
satisfy:
\begin{equation}
0= S_{\epsilon},_i(\phi^j_{\epsilon}) =
S,_i(\phi^j_{\epsilon}) + \epsilon A,_i(\phi^j_{\epsilon})
\end{equation}
To first order then, the perturbations $\delta \phi^i$ are
governed by the equation:
\begin{equation}
\label{sd}
-A,_i(\phi^k) = S,_{ij}(\phi^k) \delta \phi^j
\end{equation}
and we see that both the boundary conditions (advanced or
retarded) and any gauge fixing applies only to the
inversion of the operator $S,_{ij}(\phi^k)$ in the above
linear equation for $\delta \phi^j$ and not to the
solution of \ref{eom} for $\phi^i$.  In the case where
there are no gauge symmetries, $S,_{ij}$ is invertible and
has advanced and retarded Green's functions $G^{\pm jk}$
that  satisfy \begin{equation}
\label{inverseS}
S,_{ij} G^{\pm jk} = - \delta^k_i
\end{equation}
so that the advanced and retarded solutions to the above
equations are $\delta ^{\pm}\phi^j = G^{\pm ji}A,_i$ where
both $G^{\pm ji}$ and $A,_i$ depend on the unperturbed
solution $\phi^i$.  Since $\delta ^{\pm}B = B,_i \delta^{\pm}
\phi^i$, the Peierls bracket is just
\begin{equation}
\label{alg}
(A,B) = A,_i \tilde{G}^{ij} B,_j
\end{equation} where
\begin{equation}
\label{tildeG}
\tilde{G}^{ij} = G^{+ij} - G^{-ij}
\end{equation}

As an aside, recall that when
using the condensed notation it is important to note
that contractions $a_ib^i$ involve integrations over time
(and space in a field theory) so that this operation may
not be associative. Associativity $(a^ib^j_i)c^j
= a^i(b^j_ic^j)$ is guaranteed only
when the various spacetime integrals converge
appropriately.   Nevertheless, because they involve only
matrices  of compact (usually local) spacetime
support and Green's functions that satisfy
the proper (advanced or retarded) boundary conditions
the expressions we consider {\it do} converge in the
required fashion and the order of contractions will not be
specified.  This subtlety should nonetheless be kept in
mind as convergence should be checked whenever an
association is to be performed.

We may now take
a different approach and {\it define} the Peierls bracket
by \ref{alg} instead of \ref{Peierls}.  Since \ref{tildeG}
is defined for all $\phi^i \in {\cal H}$ for which
$S,_{ij}(\phi)$ is invertible, this new ``extended"
Peierls bracket $(,)_{\cal H}$ is defined on a much larger
space.   In practice, we will ignore any difference
between this space and ${\cal H}$ itself.

The term ``extension" is appropriate because this large
bracket has a well-defined pull back to ${\cal S}$ where
it coincides with the original Peierls bracket.  To see
this, we note that $S,_{ij}\tilde{G}^{jk} = 0$ so that we
have \begin{equation}
\label{proj}
(A + a^i S,{}_i, B+b^jS,{}_j)_{\cal H} = (A,B)_{\cal H} +
c^kS,{}_k \end{equation}
Thus, if $i: {\cal S} \rightarrow {\cal H}$ is the
inclusion map, $F$, $G$, $J$, and $K$ are sufficiently
smooth functions on ${\cal H}$ such that $F \circ i = J
\circ i$ and $G \circ i = K \circ i$, and $S,_i$ behave
like coordinate functions near ${\cal S}$  it follows that
$(F,G)_{\cal H}\circ i =  (J,K)_{\cal H} \circ i$ and that
we may consistently define a bracket $(,)_{\cal S}$ on
${\cal S}$ by \begin{equation}
\label{S pull back}
(F \circ i, G\circ i)_{\cal S} \equiv (F,G)_{\cal H} \circ i
\end{equation}
This $(,)_{\cal S}$ is just the original Peierls bracket.

As with $(,)_{\cal S}$, the larger $(,)_{\cal H}$ forms a
Lie bracket.  Antisymmetry is guaranteed since $S,_{ij}$
is  symmetric and therefore $G^{+ij}=G^{-ji}$.  The
derivation requirement is satisfied by construction and
the Jacobi identity follows by
a straightforward computation using the fact that
$\tilde{G}^{ij},_k = G^{+im}S,_{mnk}G^{+nj} -
G^{-im}S,_{mnk}G^{-nj}$ and that $S,_{mnk}$ is symmetric.

Because this extension is defined by a matrix of
second derivatives and not by a tensor field
on ${\cal H}$, this bracket is
invariant only under {\it linear} changes
of coordinates, just as was the extended Poisson bracket
defined in \cite{other}.  The extension of the Peierls
bracket is therefore not determined by $S$ and the
manifold structure of ${\cal H}$ alone but
requires a linearized structure.  Such a structure was defined
in \cite{other} to be a set of coordinate patches that
cover ${\cal H}$ and for which the transition functions
between patches
are linear.  This structure is also known as a set of linearized
coordinates and it defines a linearized manifold.
Our algebra may be defined
on any linearized manifold ${\cal H}$ by assembling
algebras defined in patches in the
manner described by appendix A of \cite{other}.

This coordinate dependence could be removed by introducing
a covariant derivative operator (;) on ${\cal H}$.  The
condition that $S_{;ij}$ be symmetric requires the
associated connection to be torsion-free.  We choose not
to use this approach here for more convenient comparison
with \cite{other} in which a coordinate independent
prescription is determined not by a covariant derivative
but by a suitable set of functions.  Such an introduction
would not significantly change our discussion as it would
only replace dependence on the linearized structure
by dependence on the choice of covariant derivative.

In the following sections, we will consider
generalizations of this extended Peierls bracket to gauge
dependent quantities. Thus, we work with algebras on
${\cal H}$ and not just on ${\cal S}$.  That our
approach is more general than a strict gauge fixing of
small disturbances around solutions will be evident from
the fact that these algebras will in general not
satisfy Eq. \ref{proj} or have well-defined pull back to
${\cal S}$.

Throughout our discussion we will draw parallels between
such generalized Peierls brackets and the various
extensions of the Poisson bracket defined in
\cite{other}.  We thus make contact with more standard
techniques through \cite{other}. As a first comparison,
Appendix \ref{equiv} shows
that the extended Peierls bracket for gauge-free systems
is identical to
the extended Poisson bracket defined by \cite{other} on
${\cal H}$ when $S$ takes the canonical form \ref{canact},
${\cal H}$ is the associated set of canonical histories,
and both algebras are defined using the same linearized
structure.

\section{Generalizations to Gauge Dependent Quantities}
\label{Gen}

The extension of the Peierls bracket to ${\cal H}$ sets
the stage for our generalization
to gauge dependent functions.  In \ref{introduce F}, we
introduce further methods of \cite{Bryce} that can be used
to define the
Peierls algebra  ${\cal A}^{GI}_L({\cal S})$ of gauge
invariant functions on ${\cal S}$ using Green's functions as was done
in IIB for gauge-free systems.
We then extend this algebra to ${\cal A}^{GI}({\cal H})$ and generalize
it to
act on gauge dependent functions in \ref{QFs}.

\subsection{Gauge Systems}
\label{introduce F}

We recall that DeWitt \cite{Bryce} has shown how
to calculate the Peierls Bracket of gauge invariants in
the presence of a gauge  symmetry in a manner similar to
\ref{alg} above.  Consider an action $S$ whose
gauge invariances are generated by $Q_{\alpha}^i$ for
$\alpha$ in some index set  $\Lambda$ that includes
(space)time labels\footnote{The set called $\Lambda$ here
was referred to as ${\cal G}^I$ in \cite{other}.}.
That is, $S$ is invariant under transformations of the
form $\delta \phi^i = \epsilon^{\alpha}Q_{\alpha}^i$
for all $\epsilon^{\alpha}$ of
compact support interior to the support of $S$.
DeWitt then
shows that the advanced and retarded effects of one
invariant $A$ on another invariant $B$ can be written in
the form \begin{equation}
\label{gauge effects}
D^{\pm}_A B =  B,_i G^{\pm ij} A,_j
\end{equation}
where $G^{\pm ij}$ are the advanced and retarded Green's
functions  that satisfy
\begin{equation}
\label{invert F}
F_{ij}G^{\pm jk} = - \delta^i_k
\end{equation}
for any operator $F_{ij}$ of the form
\begin{equation}
\label{F}
F_{ij} = S,{}_{ij} + P_{\alpha i}\eta^{\alpha \beta}
P_{\beta j} \end{equation}
Here $\eta^{\alpha \beta}$ is an arbitrary symmetric
invertible local  continuous matrix (which we will in fact
take to be ultralocal  throughout our discussion) and
$P_{\alpha i}$ are a set of one-forms on ${\cal H}$ such
that  ${\cal F}_{\alpha \beta} = P_{\alpha i} Q_{\beta}^i$
is a  non-singular differential operator.  DeWitt shows in
\cite{Bryce} that the operator $F_{ij}$ is always
invertible when ${\cal F}_{\alpha \beta}$ is invertible
and that when the source $\epsilon A$ is gauge invariant
\ref{gauge effects} - \ref{F} are
equivalent to imposing the gauge fixing conditions
$P_{\alpha i} \delta^{\pm} \phi^i =0$ on the
disturbances when calculating $\delta B$.

Greek indices will always
take values in the set $\Lambda$ while Latin indices will
take values in the set ${\cal I}$.  The combination
$P_{\alpha i}\eta^{\alpha \beta} P_{\beta j}$ is referred
to as the ``invariance breaking term" and we
will refer to $P_{\alpha i}$ and $\eta^{\alpha \beta}$
respectively as the ``invariance breaking form" and the
``invariance breaking metric" through this
``metric" has nothing to do with any spacetime metric
and may have arbitrary
signature.
Note that $\eta^{\alpha \beta}$ and $P_{\alpha i}$
need not be globally defined, nor even the invariance
breaking term as a whole, so long as $F_{ij}$ is defined
in patches that cover ${\cal S}$ in such a way that the algebras
in these patches are compatible in the sense of
appendix A of \cite{other}.  This appendix describes the
assembly of a globally defined algebra from algebras
defined in patches so long as
the Lie brackets of any two patches are identical when
acting on functions with support in the intersection of
those patches.

As with \ref{alg}, \ref{gauge effects} can be used to define a Lie
bracket $(A,B)_{\cal S} = A,_i \tilde{G}^{ij} B,_j$ of gauge
invariant functions (with spacetime support internal to that of $S$)
on ${\cal S}$.
Again, antisymmetry follows because
$F_{ij}$ is self-adjoint and again the
derivation property is immediate.  The Jacobi identity and the
pull back property (Eq. \ref{proj}) require more care but
can be verified using the fact that the algebra
is defined only on invariants and are derived
in \cite{Bryce}
by applying the useful fact:
\begin{equation} \label{useful} Q^i_{\beta}
{\cal G}^{\pm \beta \alpha} = G^{\pm ij} P_{\beta j}
\eta^{\beta \alpha} \ , \end{equation}
which holds on ${\cal S}$ and is derived in \cite{Bryce}
by applying the symmetry generators to $F_{ij}$ and using
the various definitions.
Because this bracket is defined only
on ${\cal S}$, it does not depend on the
choice of linearized structure.

That our algebra is independent of the choice of invariance
breaking term can be seen by computing the change in the advanced and
retarded Green's functions induced by a change in the
invariance breaking term:

\begin{eqnarray}
\label{vary}
\delta G^{\pm ij} &=& - G^{\pm ik} \delta F_{kl} G^{\pm lj}
\cr &=& - G^{\pm ik} \delta P_{\alpha k} \eta^{\alpha
\beta} P_{\beta l} G^{\pm lj} - G^{\pm ik} P_{\alpha k}
\delta \eta^{\alpha \beta} P_{\beta l}
G^{\pm lj} - G^{\pm ik} P_{\alpha k} \eta^{\alpha
\beta} \delta P_{\beta l} G^{\pm lj}
\end{eqnarray}
and using \ref{useful}.
We find that $\delta (A,B) = 0$
under this variation if $A$ and $B$ are both gauge
invariant.
The resulting algebra may thus be called
the extended Peierls algebra of gauge invariants.
If the gauge generators $Q^i_{\beta}$ are linear so that
$Q^i_{\beta}{},_j = 0$ then Eq. \ref{useful} in fact holds on
all of ${\cal H}$, from which it follows that the generalized
Peierls bracket may be extended in this case to ${\cal A}^{GI}_L({\cal H})$
as a Lie bracket that is independent of the choice of
invariance breaking term.
Appendix \ref{boundaries} verifies that the restriction
on the spacetime support of $A$ and $B$ is important.

\subsection{Lie Brackets of Gauge Dependent Functions}
\label{QFs}

We now generalize our bracket to act on gauge
dependent functions.  Recall that gauge invariance
of $A$ and $B$ and properties of ${\cal S}$
were required only to derive the pull back
property (Eq. \ref{proj}) and the Jacobi identity.  Therefore, if for
some choice of $F_{ij}$ the bracket  \begin{equation}
\label{def} (A,B)_{\cal H}\equiv A,_i \tilde{G}^{ij}B,_j
\end{equation}
satisfies the Jacobi identity even when $A$ and $B$ are
gauge dependent,
\ref{def} defines a Lie bracket for
all functions on the space of histories, although it may not
have well-defined pull back to ${\cal S}$.  Also, if
\ref{def} and \ref{proj} happen to hold on ${\cal S}$, then
$(,)_{\cal S}$ is a Lie bracket of functions on the space of
solutions.  Again, $A$ and
$B$ should have spacetime support interior to that of $S$,
though more general cases may be defined by limits when
such limits converge. Note that this bracket
may depend on the choice of $F_{ij}$ and may or
may not satisfy the  analogue of Eq. \ref{proj} when $A$
and $B$ are not invariants --  these are issues to be
explored in the coming sections.  In this section we
investigate the more fundamental issue of finding classes
of invariance breaking terms for which the
Jacobi identity is a consequence of \ref{def}.

We first observe that the Jacobi identity for systems
without gauge symmetries followed only from the symmetry
of $S,_{ijk}$.  Similarly then, our
generalized Peierls bracket is a Lie bracket whenever
$F_{ij}{},_k$, or equivalently, $(P_{\alpha i}
\eta^{\alpha \beta} P_{\beta j}),_k$, is symmetric.
Such invariance breaking terms will be called ``class I,"
as will the associated algebras.

To find another condition under which \ref{def}
defines a Lie bracket (though only on ${\cal S}$), we
compute the  antisymmetrized triple bracket:
$\epsilon_{\alpha \beta \gamma}
(A^{\alpha}(A^{\beta},A^{\gamma}))$   for general $A^1$,
$A^2$, $A^3$ where $\epsilon_{\alpha  \beta \gamma}$ is
the completely antisymmetric symbol with three  indices,
$\alpha, \beta, \gamma \in \{1, 2,3\}$.
Using $G^{\pm ij},_k = G^{\pm im}S,_{mnk}G^{\pm nj}$,
and Eq. \ref{useful} \cite{Bryce} arrives at \begin{eqnarray}
\label{jacobi}
\epsilon_{\alpha \beta \gamma}
(A^{\alpha},(A^{\beta},A^{\gamma}))  &=&
\epsilon_{\alpha \beta \gamma} A^{\alpha},_i A^{\beta},_j
A^{\gamma},_k \tilde{G}^{im} (G^{+jn}P_{\sigma n}{},_n
{\cal G}^{-\lambda \sigma} Q^k_{\lambda} \cr &-&
Q^j_{\sigma}  {\cal G}^{+\sigma \lambda} \eta_{\lambda
\gamma}{},_m  {\cal G}^{-\delta \gamma} Q^k_{\lambda} +
Q^j_{\sigma} {\cal G}^{+\sigma \lambda} P_{\lambda n}{},_m
G^{-kn} \cr &-& G^{-jn} P_{\sigma n}{},{}_m {\cal
G}^{+\lambda \sigma} Q^k_{\lambda} + Q^j_{\sigma} {\cal
G}^{-\sigma \lambda} \eta_{\lambda \gamma}{},m {\cal G}^{+
\delta \gamma} Q^k_{\delta} \cr &-& Q^j_{\sigma} {\cal
G}^{-\sigma \lambda} P_{\lambda n}{},_m G^{+kn})
\end{eqnarray} on ${\cal S}$ (or when $Q^i_{\beta}{},_j = 0$)
where ${\cal G}^{\pm \alpha \beta}$ are the
advanced and retarded  right Green's functions of ${\cal
F}_{\alpha \beta}$.

A common special case occurs when
${\cal F}_{\alpha \beta}$ is invertible without imposing
boundary conditions to the past or the future so that
${\cal G}^{+\alpha \beta} = {\cal G}^{- \alpha \beta} =
{\cal G}^{\alpha \beta}$.  In
this case \ref{jacobi} simplifies and vanishes
when
\begin{equation} \label{sym}
P_{\lambda n}{},_m = P_{\lambda m}{},_n
\end{equation}
Thus, the second class of algebras (class II) that we will
consider is defined by $F_{ij}$ for which ${\cal
F}_{\alpha \beta}$ is invertible without past or future
boundary conditions and $P_{\lambda m}{},_n$ is symmetric
in $n$ and $m$.  Note that this condition does not involve
the invariance breaking metric $\eta^{\alpha \beta}$ but, as
the Jacobi identity was verified using \ref{useful}, it is
guaranteed to form a Lie bracket only on ${\cal S}$ unless
$Q^i_{\beta}{},_j = 0$.

Finally, if this bracket is to be well defined on ${\cal S}$,
we must show that $(F \circ i, G \circ i)_{\cal S}$, where $F$ and
$G$ are two sufficiently smooth functions on ${\cal H}$
and $i : {\cal S} \rightarrow {\cal H}$ is the natural embedding,
does in fact depend only on $F \circ i$ and $G \circ i$.
As before, we show this by computing (on ${\cal S}$)

\begin{eqnarray}
\label{wd}
(S,_i\ ,A) = S,_{ij} \tilde G^{jk} A,_k
&=&P_{i \alpha} \eta^{\alpha \beta} P_{j \beta} (G^{+jk} - G^{-jk}) A,_k
\cr
&=& - P_{i \alpha} \tilde {\cal G}^{\beta \alpha} Q^k_{beta} A,_k
\end{eqnarray}
using Eq. \ref{useful}. This vanishes for class II so that such
generalized Peierls algebras  are well defined.

We have given no proof and
make no claim that one of these conditions is necessary
for the Jacobi identity to hold.  As a result,
it is comforting to check that there do
exist operators $F_{ij}$ such that the bracket
defined through \ref{def} does {\it not} satisfy the
Jacobi identity.  Such examples are
discussed in appendix \ref{fail} and show that neither of
the conditions that defines class II is, on its own,
sufficient to guarantee that the Jacobi identity holds.

\section{Special Cases}
\label{explore}

Because a generalized Peierls algebra is determined by a
choice of invariance breaking term, we have the potential
for a variety of distinct generalizations.  The purpose of
this section is to exhibit this variety by investigating
several invariance breaking terms and relating the results
to familiar algebras of gauge dependent
functions. In particular, \ref{gauge breaking} shows that
generalized Peierls brackets can reproduce any gauge
broken Poisson algebra defined in \cite{other} and
therefore both gauge fixed algebras and the canonical
algebra of \cite{Dirac}. \ref{Landau} shows that Feynman
and Landau ``gauge" algebras are also special cases of
generalized Peierls brackets. We have thus found a unified
description of these techniques.

\subsection{Gauge Breaking}
\label{gauge breaking}

Gauge broken algebras on ${\cal H}$ were defined in
\cite{other} by  introducing a set of locally defined
functions $P^{\alpha}[\phi^i]$ that locally foliate ${\cal
H}$ into slices ${\cal H}_{c^{\alpha}}$ on which
$P^{\alpha} = c^{\alpha}$ and another set of  locally
defined functions $\phi^a$ such that $(\phi^a,
P^{\alpha})$ is a local product structure on ${\cal H}$.
That is, $(\phi^a, P^{\alpha})$ must form coordinates in
patches such that the transitions functions preserve the
product structure defined by the separation of coordinates
into  $\phi^a$ and $P^{\alpha}$.  In addition, we ask that
the $\phi^a$ be linearized and that $\phi^a$ and
$P^{\alpha}$ be ultralocal on $M$.

If we now introduce the
inclusion map $\Phi_{c^{\alpha}}: {\cal H}_{c^{\alpha}}
\rightarrow {\cal H}$ with components
$\Phi^i_{c^{\alpha}}$, where  $\phi^i$ ranges over
$\phi^a$ and $P^{\alpha}$, then the gauge broken Poisson
bracket is defined by
\begin{equation}
\label{lift}
(A,B)_{\cal H}(p) = A,_i (\phi^i \circ \Phi_{c^{\alpha}},
\phi^j \circ \Phi_{c^{\alpha}})_{c^{\alpha}} (x) B,_j
\end{equation}
where $p = (x,c^{\alpha}) \in {\cal H}$, $x \in {\cal
H}_{c^{\alpha}}$, and $\phi^i$ ranges over $\phi^a$ and
$P^{\alpha}$. The bracket $(,)_{c^{\alpha}}$ is the
extended Poisson bracket defined by the linearized
structure of the coordinates $\psi^a = \phi^a \circ
\Phi_{c^{\alpha}}$ and the restriction $S_{c^{\alpha}} = S
\circ \Phi$ (which has no gauge freedom)
of the original action to ${\cal
H}_{c^{\alpha}}$ when $S_{c^{\alpha}}$ takes the canonical
form.  When  $S_{c^{\alpha}}$  does not take
the canonical form,  $(,)_{c^{\alpha}}$ is
defined from an extended Poisson bracket
$(,)'_{c^{\alpha}}$ by the fact that it respects certain
equations of motion that follow from $S_{c^{\alpha}}$ and
these same equations  are solved to construct the
canonical action $S_{c^{\alpha}}'$ that defines
$(,)'_{c^{\alpha}}$. By appendix \ref{equiv},
$(,)_{c^{\alpha}}'$ is just the extended Peierls algebra
on ${\cal H}_{c^{\alpha}}'$.  Since the extended Peierls
bracket on ${\cal H}_{c^{\alpha}}$ respects all equations
of motion, it then follows from appendix \ref{solving}
that, for a given linearized structure on ${\cal
H}_{c^{\alpha}}$, $(,)_{c^{\alpha}}$ is the Peierls
bracket on ${\cal H}_{c^{\alpha}}$ defined by
$S_{c^{\alpha}}$.   This Peierls bracket is built from the
Green's functions of

\begin{equation}
\label{hom}
S_{c^{\alpha};ab} = \Phi^i_{c^{\alpha};a} S,_{ij}
\Phi^j_{c^{\alpha}jb}
\end{equation}
where the semicolon denotes a derivative with respect to
the coordinates $\psi^a$ on ${\cal H}_{c_{\alpha}}$.  Our
task is now to show how the proper choice of invariance
breaking term can generate the operator \ref{hom}.

To do so, we first consider a system
described by an arbitrary action functional $S[\phi^i]$
together with an arbitrary set of cotangent vectors
$P_{\alpha i}$ locally defined on ${\cal H}$ and
ultralocal on $M$.   Note that the covectors
$P_{\alpha i}$ form a one-form that takes values in the
dual of the algebra of gauge  generators.  As such, it
annihilates some section $N$ of the tangent bundle
$T^*{\cal H}$.  Since $P_{\alpha i}$ is ultralocal on $M$,
we may choose a set $\{\pi_a^i\}$ of basis
vector fields (labelled by the index $a$) for $N$ that are
ultralocal on $M$ and defined locally on ${\cal H}$ such
that, together with a set $\{\pi_{\alpha}^i\}$ of vector
fields also ultralocal on $M$ and locally defined on
${\cal H}$ but labelled by an index $\alpha \in \Lambda$
they form a basis $\{\pi_j^i\}$ of $T^p{\cal H}$ at
each $p \in {\cal H}$.  The matrix $\pi^{-1}{}_j^i$ is
uniquely defined since the $\pi^i_j$ are ultralocal in
$M$.  We use latin indices ($j$) to run over both kinds of
basis vectors and note that the $\{\pi^i_a\}$ are local
components of a projection $\pi: T_*{\cal H} \rightarrow
N_*$ where $N_*$ is the dual of $N$.

What we would like to show is that
in the limit of uniformly large eigenvalues of
$\eta^{\alpha \beta}$,  the algebra defined by $F_{ij}$
reduces to the algebra defined through \ref{Peierls}
by the operator $S,_{ij}$ projected onto $N$ by $\pi^i_a$.
That is,
we would like to show that this limit is well defined if
\begin{equation}
\label{Sab}
S^N_{ab} \equiv \pi_a^i S,_{ij} \pi_b^j
\end{equation}
is invertible and that
the limit of $\tilde{G}^{ij}$ is given by the pull-back of
$G^{N+ab} - G^{N-ab}$ as a bilinear form on $N_*$ to a
bilinear form on $T_*{\cal H}$  through the
projection $\pi$.  Here, $G^{N\pm ab}$ are the advanced and retarded
Green's functions of $S^N_{ab}$.

We begin with the projection of
the defining equation \ref{invert F} on the left:
\begin{equation}
\pi_a^i S,_{ij} G^{\pm jk} = - \pi_a^k
\end{equation} which, after multiplying by $G^{N\pm
da}$ and expanding the identity operator on $T^*{\cal H}$
as $\delta^i_j = \pi^i_k \pi^{-1}{}_j^k = \pi^i_a
\pi^{-1}{}_j^a + \pi^i_{\alpha} \pi^{-1}{}^{\alpha}_j$,
may be rewritten as \begin{equation}
\label{G relation}
\label{g0proj} G^{N\pm bc} \pi_c^k - \pi^{-1}{}^b_j G^{\pm
jk} + G^{N\pm bc} \pi_a^i S,_{ij} \pi_{\alpha}^j
\pi^{-1}{}^{\alpha}_m G^{\pm mk} = 0 \end{equation} Note
that if we now contract this equation with
$\pi^{-1}{}_k^{\beta}$ the first term is annihilated and
the remaining terms give a linear relation between
$\pi^{-1}{}_j^bG^{\pm jk} \pi^{-1}{}_k^{\beta}$ and
$\pi^{-1}{}_j^{\alpha}G^{\pm jk} \pi^{-1}{}_k^{\beta}$
with invertible coefficients that are independent of
$\eta^{\alpha \beta}$.  It follows that these two
quantities must be  of the same order in the scale $\eta$
of the eigenvalues of $\eta^{\alpha \beta}$ for large
$\eta$.

If we project \ref{invert F} on both sides using
$\pi_{\alpha}^i$ on the left and $\pi^{-1}{}_k^{\beta}$ on
the right, the result again involves only the two
projections $\pi^{-1}{}_j^bG^{\pm jk}
\pi^{-1}{}_k^{\beta}$ and   $\pi^{-1}{}_j^{\alpha}G^{\pm
jk} \pi^{-1}{}_k^{\beta}$  of the Green's functions
$G^{\pm ij}$: \begin{eqnarray} -\delta_{\alpha}^{\beta}
&=& \pi_{\alpha}^i S,_{ij} \pi_a^j \pi^{-1}{}^a_m G^{\pm
mk} \pi^{-1}{}_k^{\beta} + \pi_{\alpha}^i S,_{ij}
\pi_{\gamma}^j  \pi^{-1}{}^{\gamma}_m G^{\pm mk}
\pi^{-1}{}_k^{\beta} \cr &+& \pi_{\alpha}^i P_{\sigma i}
\eta^{\sigma \tau} P_{\tau j} \pi^j_{\gamma}
\pi^{-1}{}_m^{\gamma} G^{\pm mk} \pi^{-1}{}_k^{\beta}
\end{eqnarray} in which the last term is the largest for
large $\eta$.  Since $\pi^i_{\alpha}$ and $P_{\sigma i}$
are ultralocal in $M$ and $\pi^i_{\alpha} \notin N$,
$\pi^i_{\alpha}P_{\sigma i}$ must be invertible.  It
follows that $\pi_{\alpha}^i P_{\sigma i} \eta^{\sigma
\tau} P_{\tau j} \pi^j_{\gamma}$ is also invertible and
that  the projected Green's functions
$\pi^{-1}{}_m^{\gamma} G^{\pm mk} \pi^{-1}{}_k^{\beta}$
are of order $1/\eta$ and vanish in the large $\eta$
limit.  By the discussion above, $\pi^{-1}{}_m^{b} G^{\pm
mk} \pi^{-1}{}_k^{\beta}$ must also vanish in this limit
as must $\pi^{-1}{}_m^{\beta} G^{\pm mk} \pi^{-1}{}_k^{a}$
since $F_{ij}$ is self-adjoint.

Returning to \ref{G relation}, consider the
 projection through $\pi^{-1}{}^{\alpha}_k$ on the right.
All that remains in the large $\eta$ limit is
\begin{equation}
\label{large eta}
G^{N\pm ba} = \pi^{-1}{}_j^b G^{\pm jk} \pi^{-1}{}^a_k
\end{equation}
Together with our results about the other
projections of $G^{\pm ij}$, \ref{large eta} implies that
$G^{\pm ij} \rightarrow \pi^i_a G^{N\pm ab} \pi^j_b$ so
that in the large $\eta$
limit the Green's functions of $F_{ij}$ become the
pull-backs through the projection $\pi$
of the Green's functions of $S^N_{ab}$ considered as a
bilinear form on $N_*$.

We have only to relate our basis $\pi^i_j$ to the
inclusion map $\Phi_{c^{\alpha}}$ of \ref{hom} and verify
\ref{lift} for an  appropriate generalized Peierls algebra
and our task will be complete. To do so, recall that local
functions $P^{\alpha}$ were introduced to define the gauge
broken algebra and that the set $(\phi^a,P^{\alpha})$
formed local coordinates on ${\cal H}$.  As a result,
$P^{\alpha},_{a} = 0.$ Now, since
$\Phi^{\beta}_{c^{\alpha}} = c^{\beta}$, we have
$\Phi^{\beta}_{c^{\alpha};a} = 0$ and $P^{\alpha},_i
\Phi^i_{c^{\alpha};a}  = 0$.  Thus, if we choose the
invariance breaking form to be $P_{\alpha i} \equiv
(\gamma_{\alpha \beta} P^{\beta}),_i$ where
$\gamma_{\alpha \beta}$ is some locally defined
nonsingular ultralocal matrix that does not depend on the
$\phi^a$, then $\pi_a^i(p) = \Phi^i_{c_{\alpha};a}(x)$ for
$p = (x,c_{\alpha})$, $x \in {\cal H}_{c_{\alpha}}$ form
an appropriate basis for $N$.  We note that such
$\gamma_{\alpha \beta}$ exist whenever $(\phi^a,
P^{\alpha})$ form the required local product structure.
The algebra defined by $(A,B)_{{\cal H}_{c^{\alpha}}} = A,_a
\tilde{G}^{Nab} B,_b$ is then identical to the extended
Poisson bracket $(,)_{c^{\alpha}}$.  Equation \ref{lift}
follows since  $\pi^{-1}{}_m^{b} G^{\pm mk}
\pi^{-1}{}_k^{\beta}$, $\pi^{-1}{}_m^{\alpha} G^{\pm mk}
\pi^{-1}{}_k^{a}$, and $\pi^{-1}{}_m^{\alpha} G^{\pm mk}
\pi^{-1}{}_k^{\beta}$ all vanish in the large $\eta$ limit.

We have now shown that the invariance breaking term may be
chosen such that the generalized Peierls bracket coincides
with any gauge broken Poisson bracket, at least in the
limit of large $\eta$. That this may be done within class
I follows from the fact that $P^{\alpha},_{ij} = 0$ in the
coordinates chosen.  Thus, setting $\eta^{\alpha \beta}
= \eta \gamma^{\alpha \beta}$ where $\eta$ is some constant that
we take to infinity and $\gamma^{\alpha \beta}$ is the inverse of
$\gamma_{\beta \delta}$, we find that
$P_{\alpha i} \eta^{\alpha \beta}P_{\beta j} =
P^{\alpha},_i \eta \gamma_{\alpha \beta} P^{\beta},_j$ so that
$(P_{\alpha i} \eta^{\alpha \beta} P_{\beta j}),_k = 0$ and
the invariance breaking term is class I.
By the discussion of \cite{other},
this means that the generalized Peierls bracket can
reproduce both gauge fixed algebras and the canonical
algebra of \cite{Dirac} on ${\cal H}$.

We note also that gauge broken
algebras based on canonical gauge fixing are type II since
$P_{\alpha i},{}_{j} = P_{\alpha j},{}_i$ and
$\gamma_{\alpha \beta}  P^{\beta}$ and ${\cal F}_{\alpha
\beta} = Q^i_{\beta} (\gamma_{\alpha \sigma}
P^{\sigma}),_i$ are ultralocal.
However,  the canonical algebra of
\cite{Dirac} is only of type I, since it is reproduced by
$\gamma_{\alpha \beta} = \delta_{\alpha \beta}$,
$P^{\alpha} = \lambda^{\alpha}$, and $\eta^{\alpha \beta}
= \delta^{\alpha \beta}$ and since
$Q^i_{\beta}(\delta_{\alpha
\sigma} \lambda^{\sigma}),_i$ is not ultralocal.  Finally, we
note that the choice of $P_{\alpha i}$ ultralocal in $M$
was not strictly necessary in the argument above but
served to simplify the discussion.

If the above invariance breaking term is in class II and we are
working on ${\cal S}$ or with linear gauge transformations,  we
need not even take the large $\eta$ limit to attain this result.
To see this,
consider the change induced in the generalized Peierls algebra
by a change $\delta \eta^{\alpha \beta}$ of this metric.  In particular,
we evaluate the derivative:

\begin{equation}
{{\delta G^{\pm ij}} \over {\delta \eta^{\alpha \beta}}}
= \eta_{\alpha \gamma} {\cal G}^{\pm \gamma \sigma}
Q^i_{\sigma} Q^i_{\lambda} {\cal G}^{\pm \kappa \lambda}
\eta_{\beta \kappa} \end{equation}
using \ref{vary} and
\ref{useful}.  Thus, ${\displaystyle {{\delta
\tilde{G}^{ij}} \over  {\delta \eta^{\alpha \beta}}}}$
vanishes on ${\cal S}$ for class II invariance breaking terms when
\ref{useful} holds since ${\cal G}^{+ \alpha
\beta} = {\cal G}^{- \alpha \beta}$ and the limiting algebra is given
by any finite ${\eta}^{\alpha \beta}$.

\subsection{Feynman and Landau Gauge}
\label{Landau}

Another common technique \cite{IZ} in quantum field theory
is to remove the gauge invariance of some action $S$ by
adding to it a term  $\Delta S$ which is a local quadratic
form in the fundamental fields $\phi^i$.  One such term is
added for each gauge invariance so that the full
modification may be written \begin{equation}
\Delta S = \gamma P_{\alpha i} \phi^i \eta^{\alpha \beta}
P_{\beta i} \phi^i \end{equation}
where $P_{\alpha i}$ and $\eta^{\alpha \beta}$ are field
independent.  With $\gamma = 1$, this is a generalization
of ``Feynman gauge" while a generalized ``Landau gauge"
arises in the limit $\gamma \rightarrow \infty$.  Note
that, by appendix \ref{equiv}, the extended Poisson
algebra that follows from $S + \Delta S$ is the
generalized Peierls algebra of type I for the original
action $S$ and the invariance breaking term $\Delta
S,_{ij}$.  Interestingly, section \ref{Gen} then
guarantees that the  algebra of gauge invariants defined
by $S$ is identical to the algebra of those same functions
defined by $S+ \Delta S$.

\section{Pull backs of $(,)_{\cal H}$}
\label{pull backs}

As in \cite{other}, it is of interest to consider pull
backs of $(,)_{\cal H}$ to spaces of partial solutions.
Such pull backs have a larger coordinate invariance than
$(,)_{\cal H}$ and are interesting for quantization both
because, since the equations of motion hold, they
lead naturally to Heisenberg picture formulations and
because representations of a commutator algebra based on
$(,)_{\cal H}$ tend to be reducible to representations
based on such a pull back (see \cite{other}). In
\ref{projections} we identify subspaces ${\cal A} \subset
{\cal H}$ to which this pull back is well-defined and in
\ref{constraints} we study the properties of the pulled
back algebras.

\subsection{Allowed Spaces}
\label{projections}

We would like to address the question: ``To which spaces ${\cal A}
\subset {\cal H}$ of partial solutions
does $(,)_{\cal H}$
have a well-defined
pull back?" For such a space, the addition of
arbitrary quantities to $A$ and $B$ that vanish on ${\cal
A}$ modifies $(A,B)_{\cal H}$
only by a term that also vanishes on ${\cal A}$.  Thus, ${\cal A}$
is characterized by a set $\{c^i_dS,_i=0\}$ of combinations
of the equations of motion
$\{S,_i=0\}$ for which $(A + a^dc_d^i S,_i, B+ b^dc_d^iS,_i)_{\cal H}
= (A,B)_{\cal H} + q^dc_d^iS,_i$ for
all $A$, $B$, $a^d$, and $b^d$ and some $q^d$ that depends on $A$, $B$,
$a^d$, and $b^d$.
In particular, $(A,c^i_dS,_i) = q^f_dc_f^iS,_i$.

Evaluating the bracket of an equation of motion, we find
\begin{equation}
\label{algeom}
(A, \ S,_k)_{\cal H}
= A,_j \tilde{G}^{jk} S,_{jk} = - A,_j \tilde{G}^{jk}P_{\alpha
k} \eta^{\alpha \beta} P_{\beta k}
\end{equation}

In general, \ref{algeom} will vanish for some
equations of motion $S,_k$ but not for others.  For
example, we have $(A, S,_ka^k)_{\cal H} = A,_i
\tilde{G}^{ij}a,_j^kS,_k$ when $a^k P_{\alpha k} = 0$.
Any algebra may thus be consistently pulled back to
any subspace defined by $c_d^iS,_i = 0$ for some
$\{c^i_d\}$ such that $c^i_dP_{\alpha i} = 0$ and
$c^i_{d,j} = K^b_{dj}c^i_d$ where $K_{dj}^b$ and $c_d^i$
have compact support.

Let us also recall Eq. \ref{wd}:
\begin{equation}
\label{bracket with S}
(A, \ S,_k)_{\cal H} = - A,_j Q^j_{\alpha}
\tilde{\cal G}^{\alpha \beta} P_{\beta k}
\end{equation}
which holds on ${\cal S}$ or when $Q^i_{\beta}{},_j = 0$ and which
vanishes when $\tilde{\cal G}^{\alpha \beta} = 0$.
This shows that any algebra that is both class I and class II has a
well-defined pull back to ${\cal S}$.
In particular, this is true of any gauge broken algebra based on
canonical gauge fixing.

\subsection{Properties of Pulled Back Algebras}
\label{constraints}

As hinted above, the properties of a pull back of a
class I generalized Peierls algebra depend on whether or not the
algebra is also in class II.  When it is, the algebra may be
pulled back to ${\cal S}$  where it depends only
on the invariance breaking term and the manifold structure of ${\cal H}$.
Algebras not of class II can only be pulled back to a
larger space of partial solutions, which we will call
${\cal E}$.

On ${\cal S}$ class II algebras have
locally nontrivially centers.  That is, the patches on
${\cal S}$ may be chosen such that the local
algebra of each patch has nontrivial center. This follows
since $P_{\alpha i},_j = P_{\alpha j},_i$ so that locally
$P_{\alpha i} = P_{\alpha},_i$ for some $P_{\alpha}$.  We
then have $(A,P_{\alpha}) = A,_i \tilde{G}^{ij}P_{\alpha
j} = 0$ as a  result of \ref{useful} since ${\cal G}^{+
\alpha \beta} =  {\cal G}^{- \alpha \beta}$.  By
\ref{gauge breaking}, the same is true on ${\cal H}$ for any algebra
defined by a large $\eta$ limit.  Class II algebras thus
have many of the same properties as gauge broken algebras
based on canonical gauge fixing  and quantization may
proceed by imposing all of the equations of motion as
operator equations with the results similar to V B of
\cite{other}.

On the other hand, the situation for algebras not of type
II is similar to the canonical case of \cite{Dirac}.
This is no surprise after \ref{gauge breaking} where we
saw that ${\cal A}_H (\Gamma)$ is a class I generalized
Peierls algebra that is not in class II.

We now work in the space ${\cal E}$ containing
${\cal S}$ and on which those equations of motions that
$(,)_{\cal H}$ respects vanish identically.  In the
Peierls equivalent of the
canonical approach (in which $P_{\alpha i}$ vanishes when the
index $i$ does not correspond to a Lagrange multiplier)
these are the equations of motion
generated by the Hamiltonian and the remaining equations
of motion are the constraints.  In general then, we
refer to those equations of motions not respected by
$(,)_{\cal E}$ as ``generalized constraints."

For quantization, the constraints should form a first
class set.   A set $\{\phi_{\alpha}\}$
of constraints is called first class when i) the bracket
of two constraints is a linear combination of
constraints:  $(\phi_{\alpha},\phi_{\beta}) =
C^{\gamma}_{\alpha \beta} \phi_{\gamma}$ and ii) when the
set of constraints does not generate further constraints
by evolution.  Here  $C^{\gamma}_{\alpha \beta}$ are
functions on ${\cal E}$.

Because our constraints carry an index that contains a
(space)time label, the set $\{\phi_{\alpha}\}$ already
contains  the evolved versions of
any constraint.  The second condition is then trivially
satisfied.  While it is quite another
issue whether or not the set of generalized constraints is equivalent to
a subset for which the index $\alpha$ lies in some
single Cauchy surface, we
have no need for such an assumption.

To show that the first condition is always
satisfied as well, we note that
\break since $(S,_iQ^i_{\alpha})_,j = 0$,
\begin{eqnarray}
\label{fc}
(S,_i\ , \ S,_k)_{\cal H} &=& -P_{\alpha i} \eta^{\alpha \beta} P_{\beta j}
\tilde{G}^{jm} S,_{km} \cr
&=& - P_{\alpha i} \tilde{\cal G}^{\alpha \beta} Q^m_{\beta} S,_{km}
- {\cal G}^{+ \alpha \beta}Q^j_{\beta}{},_n S,_jG^{+nm}S,_{km}
+ {\cal G}^{- \alpha \beta}Q^j_{\beta}{},_n S,_jG^{-nm}S,_{km}
\cr
&=&  P_{\alpha i} \tilde{\cal G}^{\alpha \beta} Q^m_{\beta}{},_k S,_{m}
- {\cal G}^{+ \alpha \beta}Q^j_{\beta}{},_n S,_jG^{+nm}S,_{km}
+ {\cal G}^{- \alpha \beta}Q^j_{\beta}{},_n S,_jG^{-nm}S,_{km}
\end{eqnarray}
and the algebra of equations of motion on ${\cal H}$ is first
class.  Since all equations of motion that are not
constraints vanish on ${\cal E}$, the algebra of
generalized constraints is the pull back to ${\cal E}$ of
this algebra on ${\cal H}$.  It follows that the
constraint algebra is first class as well.  This is
related to the fact derived in appendix \ref{equiv} that
when all constraints are second class the extended Peierls
algebra produces not the extended Poisson algebra, but
the extended Dirac algebra.  Thus, any extended
Peierls bracket can be used to define a quantum theory by
following Dirac's procedure \cite{Dirac} in which the
constraints are imposed as conditions to select
``physical" states.

Another similarity with \cite{Dirac} is that when $S,_k$
is a constraint, \ref{bracket with S} \ shows that it
generates the transformation $\delta \phi^i = Q^i_{\alpha}
\xi^{\alpha}_k$ where  $\xi^{\alpha}_k = \tilde{\cal
G}^{\alpha \beta} P_{\beta k}$ on ${\cal S}$ (or on ${\cal E}$,
up to terms proportional to the constraints).  This is not a gauge
transformation, however, since $\xi^{\alpha}_i$ does not
have compact support.  Instead, when $P_{\alpha i}
= P_{\alpha},{}_i$, since $P_{\alpha i}
Q^i_{\beta} \xi^{\beta}_k = {\cal F}_{\alpha \beta}
\tilde{\cal G}^{\beta \gamma} P_{\gamma k} = 0$ this
transformation
corresponds to one of the symmetries that would remain if
the gauge conditions $P_{\alpha} (\phi^i) = 0$ were
imposed.  This is just what was found for the extended
canonical Poisson bracket in \cite{other} and the rest of
section V A of \cite{other} applies to this case as well.

\section{Discussion}
\label{Dis}

We have seen that the machinery of \cite{Bryce}
can be used to first extend the Peierls bracket from the
space of solutions to the space of histories and then to
generalize it to act on gauge dependent functions.  This
generalization depends on the choice of an ``invariance
breaking term" which must be chosen in the proper way so
that the generalized Peierls algebra is a Lie algebra.
Two interesting classes of invariance breaking terms
were identified, one that defines a Lie algebra on ${\cal A}_L({\cal H})$
and one that is guaranteed to be a Lie algebra only on
${\cal A}_L({\cal S})$.

Our generalized Peierls bracket includes the extended
Poisson bracket of \cite{other} as a special case
so that \cite{other} provides a comparison with more
familiar methods.  Because it includes Landau and Feynman
gauge as well, the generalized Peierls algebra provides a
unified descriptions of the conventional algebras of gauge
dependent functions.  Generalized Peierls algebras
resemble the constrained canonical algebras of
\cite{Dirac} and lead to a set of ``generalized
constraints." Because these are always first class,
quantization may proceed by analogy with \cite{Dirac}.

Thus, we have described a large class of classical
(commuting) *-Lie algebras of complex functions on ${\cal
H}$, ${\cal S}$, and ${\cal E}$.  They are constructed
without performing a 3+1 decomposition so that their
covariance is manifest and depends only on the covariance
of the invariance breaking term.  No gauge fixing or
reduction is needed and, because the algebra is defined on
functions of histories, a Heisenberg picture is the
natural quantization.

\acknowledgements
This work was partially supported by a National Science
Foundation Graduate Fellowship, by NSF grants
PHY90-05790 and PHY93-96246, and by research funds provided by Syracuse
University.  Special thanks go to the referee for Annals of Physics
for discovering a mistake in an earlier version.
The author would also like to express his thanks
to Carlos Ordo\~nez, Josep Pons, and the Syracuse
relativity group for their support and to Bryce DeWitt,
without whom this project would never have begun.

\appendix

\section{Equivalence of Peierls and Dirac brackets on the
space of  Histories}
\label{equiv}

Peierls original paper \cite{Peierls} shows that, for an
unconstrained system, the Peierls bracket is mapped to the
Poisson bracket under any map $e_t: {\cal S} \rightarrow
\Gamma$ that evaluates the canonical fields at some time
$t$.  In fact, when there are no constraints or gauge
symmetries, the extended Peierls bracket is identical to
the extended Poisson bracket and when only second class
constraints are present, it yields the extended Dirac
bracket of \cite{other}.

Recall that for a system
described over a time interval $I$ by the action
\begin{equation}
\label{canact}
S =
\int_I dt(\case{1}{2}\Omega^{-1}_{AB} z^A(t) \dot{z}^B(t)
- H(t) - \lambda^a \xi_a(t)) \end{equation}
for some
field and time independent invertible  antisymmetric
matrix $\Omega^{AB}$, $H(t) = H(z^A(t),t)$, $\xi_a(t) =
\xi(z^A(t),t)$ for $a$ in some index set ${\cal T}$,
and some choice $z^A(t)$ of linearized coordinates
on a phase space $\Gamma$, the extension of the Poisson
bracket in \cite{other} to ${\cal H} \subset
\Gamma^I \times {\bf L}^I$, where ${\bf L}$ is the range
of $\lambda^a$, is uniquely determined by the fact that it
is a Lie bracket, that it respects the equations of motion
$\{S,_{(A,t)} = 0\}$,  and that $(z^A(t),z^B(t))_{\cal H}
= \Sigma^{AB} \equiv \Omega^{AB} - \Omega^{AB} \xi_{a|C}
(\Delta^{-1})^{ab} \xi_{b|D} \Omega^{DB}$. The matrix
$\Delta_{ab} = \xi_{a|A} \Omega^{AB} \xi_{b|B}$ is
invertible  since the constraints are second class. Here,
$_{|A}$ denotes a derivative with respect to the canonical
coordinate $z^A$ on $\Gamma$ and we have separated the
time label $t$  from the other labels $A$ and $a$ so that
our usual  condensed index is $i = (A,t)$ or $i = (a,t)$.

We have already seen
that the extended
Peierls bracket is a Lie bracket and that it respects the
same equations of motion when built using the same
linearized structure.  Thus, we now
compute $(z^A(t),z^B(t))_{\cal H}$ by inverting the
operator $S,_{ij}$:

\begin{mathletters}
\begin{equation}
S,_{(a,t_1)(b,t_2)} = 0
\end{equation}
\begin{equation}
S,_{(a,t_1)(B,t_2)} = \xi_{a|B}(t_1) \delta(t_1 - t_2)
\end{equation}
\begin{equation}
S,_{(A,t_1)(b,t_2)} = \xi_{b|A}(t_1) \delta(t_1 - t_2)
\end{equation}
\begin{equation}
S,_{(A,t_1)(B,t_2)} = \Omega^{-1}_{AB} \delta'(t_1-t_2) - H_{\xi|AB}(t_1)
\delta(t_1-t_2)
\end{equation}
\end{mathletters}
where the $'$ denotes a derivative with respect to its
argument and $H_{\xi}(t) = H(t) - \lambda^a(t) \xi_a(t)$.
{}From the two equations \begin{equation}
S,_{(a,t_1) i} G^{i (B,t_2)} = 0 \qquad \text{and} \qquad
S,_{(A,t_1) i} G^{i (B,t_2)} = - \delta^B_A \delta(t_1 -
t_2) \end{equation}
it follows that the Green's functions $G^{(A,t_1)(B,t_2)}$
are: \begin{equation}
G^{(A,t_1)(B,t_2)} = \pm \Sigma^{AC} {\cal P}
\exp[\int_{t_2}^{t_1}dt \ {\bf Q}(t)]^B_C
\theta[\mp(t_1-t_2)] \end{equation}
where $\theta$ is the usual step-function, ${\cal P}$
denotes path ordering, and
\begin{equation}
{\bf Q}^B_C = \Sigma^{BD} H_{\xi|DC} -
\Omega^{BD} \xi_{b|D} \Delta^{-1ba} {{\partial} \over
{\partial t}} (\xi_{a|B})]
\end{equation}
Finally, we have
$(z^A(t),z^B(t)) = \tilde{G}^{(A,t)(B,t)} = \Sigma^{AB}$
and it follows from \cite{other} that
the extended Dirac and Peierls brackets agree on ${\cal
H}$.

As discussed in section IV of \cite{other}, spaces ${\cal
L}$ of Lagrangian histories can typically be embedded as
spaces of partial solutions in spaces of canonical
histories.  The Dirac bracket on ${\cal L}$ is defined in
\cite{other} by pull back through this embedding map,
while appendix \ref{solving} shows that the same pull back
takes the extended Peierls bracket on ${\cal H}$ to the
extended Peierls bracket on ${\cal L}$.  It follows that
the extended Dirac bracket and extended Peierls brackets
are also identical on typical spaces of Lagrangian
histories.

\section{Gauge Invariants with wider Support}
\label{boundaries}

In this appendix we verify that the restriction of the
Peierls algebra to gauge invariant functions with compact
spacetime support {\it interior} to that of $S$ is
essential.  Specifically, we give an example for which the
generalized Peierls bracket of gauge invariants whose
support is not interior to that of $S$ depends on the
invariance breaking term in \ref{F}.

Consider a free relativistic particle described by the
action: \begin{equation}
S = \case{1}{2} \int^{t_2}_{t_1} ({{\dot{x}^2} \over N} -
Nm^2) \end{equation}
and two invariance breaking terms: one that leads to the
canonical algebra:

\begin{eqnarray}
(x^{\mu}(t),x^{\nu}(t))_c = 0, \qquad (x^{\mu}(t),
\dot{x}^{\nu}(t))_c = N(t), \cr (\dot{x}^i(t),\dot{x}^j(t))_c = 0,
\quad (A,
N(t))_c = 0 \end{eqnarray}
for any function $A$ on ${\cal E}$, and one that leads to
the deparameterized algebra:
\begin{eqnarray}
(x^0(t),A)_d = 0, \qquad (N(t),A)_d =
({{\sqrt{-\dot{x}^2(t)}} \over m}, A)_d \cr
(x^i(t),x^j(t))_d = 0 \qquad (x^i(t), {{\dot{x}^j(t)}
\over  {\sqrt{-\dot{x}^2(t)}}})_d
= \delta^{ij}, \qquad (\dot{x}^i(t),\dot{x}^j(t))_d = 0
\end{eqnarray}
for any function $A$ on ${\cal S}$.
Here, $i,j \in \{1,2,3\}$ and $\mu, \nu \in \{0,1,2,3\}$.  That such
terms exist are guaranteed by section \ref{explore} and
Appendix  \ref{solving}.

We now note that any quantity of the form
$\int^{t_2}_{t_1} A(t) dt$ where $A(t)$ is a scalar under
time reparameterizations is gauge invariant since gauge
transformations must vanish on the boundary.  We then
compute the bracket of \begin{equation}
T = \int^{t_2}_{t_1} N(t)dt \qquad \text{and} \qquad X^{\mu} =
\int^{t_2}_{t_1} N(t) x^{\mu}(t) dt
\end{equation}
in each algebra.  Clearly, $(T,X^{\mu})_c = 0$ and
$(T,X^0)_d = 0$.  However,
\begin{eqnarray}
(T,X^i)_d &=& T \int^{t_2}_{t_1} ({\dot{x}^0 \over
{\sqrt{m^2 + p^2(t_1)}}}, x^i(t))_d \cr &=& TX^0
{{p^i} \over {(m^2 + p^2)^{3/2}}} \neq 0 \end{eqnarray}
where $p^i(t) = \dot{x}^i(t)/\sqrt{-\dot{x}^2(t)}$ and the
algebras disagree.

A more careful
treatment of $F_{ij}$ and $\tilde{G}^{ij}$ near the past
and future boundaries than that of \cite{Bryce}
would clarify the standing of such invariants on.
However, such a treatment is complicated by the fact
that, for most systems, gauge transformations must vanish
on the boundary, even when the proper boundary terms are
included in the action.
This means that gauge transformations near the
boundary cannot be described by generators $Q^i_{\alpha}$
in quite the same way as in the interior.  The status of
$Q^i_{\alpha}$ when $i$ lies on a boundary is unclear, so
that the methods of \cite{Bryce} cannot be used directly.

Note that such difficulties never arise when comparing
algebras on ${\cal S}$.  This is because, using the
equations of motion, any gauge invariant can be written
in terms of initial data on a single Cauchy surface.
Its
algebraic properties are then determined by a gauge
invariant function with compact spacetime support
interior to the support of $S$.  It is only on ${\cal E}$
and ${\cal H}$
that each instant of time introduces genuinely new
operators, such as $N(t)$ in this example.  These lead to
gauge invariants on ${\cal E}$ like $T$ and $X^{\mu}$ that
are not determined by the initial data.

\section{Failures of the Jacobi Identity}
\label{fail}

This appendix contains two examples of operators $F_{ij}$
such that the brackets they would
define though Eq. \ref{def} do not satisfy the Jacobi
identity.  We will of course choose these $F_{ij}$
carefully so that they do not fall into either of the two
classes for which the Jacobi identity is guaranteed.  In
particular, for neither case is $F_{ij},_k$ symmetric and
we will choose one $F_{ij}$ that cannot be
written in the form \ref{F} with a  $P_{\alpha i}$ that
satisfies Eq. \ref{sym} but such that ${\cal F}_{\alpha
\beta}$ {\it is} invertible without past or future
boundary conditions and one $F_{ij}$ for which $P_{\alpha
i}$ {\it does} satisfy \ref{sym} but for which the
advanced and retarded Green's functions of ${\cal
F}_{\alpha \beta}$ are distinct.  Thus, we show that
neither condition that defines  class II is sufficient by
itself to derive the Jacobi identity.

For the first example, consider a nonrelativistic
particle in ${\cal R}^3$ with one component of
its momentum constrained to vanish and a
corresponding translational gauge symmetry.  In an
appropriate coordinate system, the canonical action
for this system takes the form: \begin{equation}
\label{ex1} S= \int dt [p_i\dot{x}^i - p^ip_i/2 - \lambda
(p_1 +p_2 +p_3)] \end{equation} where $\lambda$ is a
Lagrange multiplier and $i \in \{1,2,3\}$. Since there is
only one gauge symmetry, the invariance parameter $\alpha$
is a time parameter.  The symmetry generators are
\begin{equation} Q^{x^i(t_1)}_{t_2} = - \delta(t_1 -
t_2), \quad Q^{p_i(t_1)}_{t_2} = 0, \quad \text{and} \quad
Q^{\lambda(t_1)}_{t_2} = -\delta '  (t_1-t_2)
\end{equation}
where $'$ denotes a derivative with respect to the
argument.  If we choose $\eta^{t_1t_2} = \gamma
\delta(t_1-t_2)$, $P_{\lambda(t_1) t_2} = 0 = P_{p_i(t_1)
t_2}$, and \begin{equation}  P_{x^1(t_1) t_2} = x^2(t_1)
\delta(t_1-t_2), \  P_{x^2(t_1) t_2} = x^3(t_1)
\delta(t_1-t_2), \ P_{x^3(t_1) t_2} = x^1(t_1)
\delta(t_1-t_2) \end{equation}
then ${\cal F}_{t_1 t_2} = - (x^1 + x^2 + x^3)
\delta(t_1-t_2)$ is invertible without past or future
boundary conditions.   A short calculation shows that the
corresponding operator $F_{ij}$ cannot be written in the
form \ref{F} with a $P_{\alpha i}$ that satisfies
\ref{sym}.  A rather long calculation shows that $(x^1 +
x^2 + x^3)^5 \epsilon_{ijk}
(x^i(t_i),(x^j(t_j),x^k(t_k))$ when evaluated at
$\dot{x}^i = p^i = 0$, is a polynomial in $x^1,$ $x^2,$
$x^3$, $t_1$, $t_2$, $t_3$ with coefficient $-3$ for the
$(t_1)^2(x^2)^4$ term so that $\epsilon_{ijk}
(x^i(t_i),(x^j(t_j),x^k(t_k)) \neq 0$ and this  bracket
does not satisfy the Jacobi identity on either ${\cal S}$
or ${\cal H}$.

For the second example, we describe the relativistic
free particle by the action $S = \int dt
\sqrt{(-\dot{x}^2)}$ so that the invariance generators are
$Q^{(\mu,t_1)}_{t_2} = - \dot{x}^{\mu} \delta(t_1-t_2)$,
which generate time reparameterizations.  If we choose
$\eta^{t_1t_2} = {{m \gamma} \over {\sqrt{-\dot{x}^2}}}
\delta(t_1-t_2)$ and \begin{equation}
P_{t_2(\mu,t_1)} = - (\sqrt {- \dot{x}(t_2)^2})
,_{x^{\mu}(t_1)} = - {{\dot{x}^{\mu}} \over
{\sqrt{-\dot{x}^2}}} \delta'(t_1-t_2)
\end{equation}
then ${\cal F}_{t_1
t_2} = -\case{\partial}{\partial t_1} (\delta(t_1-t_2)
\sqrt {-\dot{x}(t_2)^2})$ is invertible.

A much shorter computation than for the first example
shows that \begin{eqnarray}
((x^{\mu}(t),x^{\nu}(t')),x^{\lambda}(t'')) &=&
\int_t^{t'} ds \big[ {{\dot{x}^{\lambda}(s) \eta^{\mu
\nu}} \over \gamma} + {{\gamma + 1} \over \gamma}
(\eta^{\mu \lambda} \dot{x}^{\nu}(s) + \eta^{\lambda \nu}
\dot{x}^{\mu}(s)) \cr &-& {{2\gamma^2 + \gamma +1} \over
{\gamma^2}} \dot{x}^{\lambda}(s) \dot{x}^{\mu}(s)
\dot{x}^{\nu}(s) \big] \end{eqnarray}
so that when $\dot{x}^{1} =1$, $\dot{x}^{2} = 1$
we have
\begin{equation}
\sum_{i,j,k \in \{1,2,3\}}
\epsilon^{ijk}
((x^{\mu_i}(t_i),x^{\mu_j}(t_j)),x^{\mu_k}(t_k))
= (t_1 - t_2) \neq 0
\end{equation}
for $\mu_1 = \mu_2 = 1,$ $\mu_3 = 2$ and this bracket
does not satisfy the Jacobi identity either.

\section{Solving Equations of Motion}
\label{solving}

In this appendix we
show that, modulo one assumption, the result of a
well-defined pull back of a generalized
Peierls algebra to a space $\hat{\cal S}$ of partial
solutions  is another generalized Peierls algebra.
Specifically,
consider such a space $\hat{\cal S}$ and the inclusion map $I:\hat{\cal S}
\rightarrow {\cal H}$.  We again assume that we have a
local product structure such that the coordinates $\phi^i$
on ${\cal H}$ separate
into two classes: $\{\phi^a\}$ and $\{\phi^z\}$ such that
the pull backs $\phi^a \circ I$ to $\hat{\cal S}$ form
coordinates on $\hat{\cal S}$ while the equations of
motion $S,_z=0$ that follow from the second set are
identically satisfied on $\hat{\cal S}$: $S,_z \circ I
\equiv 0$.  We will label these classes with indices  from
opposite ends of the alphabet.

Let $\hat{S}$ be the pull back $S \circ I$.  The
variations of $\hat{S}$ and $S$ are related by:
$\hat{S}_{;a} = (S,_i \circ I) I^i_{;a}$ where the
semicolons denote derivatives with respect to $\phi^a
\circ I$ on $\hat{\cal S}$.  The second derivatives are
related by: \begin{equation} \label{second var}
\hat{S}_{;ab} = (S,_{ij} \circ I) I^i_{;a} I^j_{;b}
\end{equation}
since $I^a_{;b} = \delta^a_b$ so that $I^a_{;bc} = 0 =
I^a_{;bz}$ and since $S,_z \circ I = 0$.

Now, suppose that some invariance breaking form $P_{\alpha
i}$ was introduced on ${\cal H}$.  We define an invariance
breaking form on $\hat{\cal S}$ by pull back:
$\hat{P}_{\alpha a} \equiv (P_{\alpha i} \circ I)
I^i_{;a}$ and similarly $\eta^{\alpha \beta} \equiv
\eta^{\alpha \beta} \circ I$.  We now have an operator
$\hat{F}_{ab}$ that is the pull back of our operator
$F_{ij}$ on ${\cal H}$: \begin{equation} \label{pull back
F} \hat{F}_{ab} = (F_{ij} \circ I) I^i_{;a} I^j_{;b}
\end{equation}

Since $I$ is an embedding, we can choose some matrix
$I^a_i$ on $\hat{\cal S}$
such that $I^a_i I^i_{;b} = \delta^a_b$.  We then note
that $\hat{G}^{\pm ab} = I^a_i G^{\pm ij} I^b_j$ are
Green's functions of $F_{ab}$ and that the Peierls algebra
they define is the pull back of $(,)_{\cal H}$ since
$(A,B)_{\cal H}\circ I  = (A,_i \tilde{G}^{ij}
B,_j) \circ I = (A,_i \circ I) I^i_{;a} (\hat{G}^{+ab}
- \hat{G}^{-ab}) I^j_{;b} (B,_j \circ I) = (A \circ
I)_{;a} (\hat{G}^{+ab}  - \hat{G}^{-ab}) (B \circ I)_{;b}$.

It follows that
the pull back of the generalized Peierls algebra to
$\hat{\cal S}$ is another generalized Peierls algebra.  In
particular, it is the one that results from pulling back
the action and the invariance breaking form to $\hat{\cal
S}$.

\end{document}